\documentclass[english]{sbrt}
\usepackage[english]{babel}
\usepackage[utf8]{inputenc}

\usepackage{graphicx}
\usepackage{amssymb}
\usepackage{algorithmic}
\usepackage{algorithm}
\usepackage{amsmath,amssymb}
\usepackage{xcolor}
\usepackage{amsmath}
\usepackage{tikz}
\usetikzlibrary{tikzmark}

\DeclareSymbolFont{slant}{OT1}{\familydefault}{m}{sl}
\SetSymbolFont{slant}{bold}{OT1}{\familydefault}{bx}{sl}
\DeclareSymbolFontAlphabet{\mathsl}{slant}
\begin{document}

\title{Simultaneous Direct and Indirect Channel Estimation for RIS-Assisted MIMO Communications}

\author{Daniel V. C. de Oliveira, Daniel C. Alcantara,  Fazal.E Asim, André L. F. de Almeida, and Gábor Fodor
\thanks{Daniel V. C. de Oliveira, Daniel C. Alcantara,  Fazal Asim, and André L. F. de Almeida are with Grupo de Pesquisa em Telecomunicações Sem Fio (GTEL), Departamento de Engenharia de Teleinformática, Universidade Federal do Ceará, Fortaleza-CE. E-mails: danielvctor@alu.ufc.br, fazalasim@gtel.ufc.br, danielchaves@alu.ufc.br, andre@gtel.ufc.br. Gabor Fodor is with Ericsson Research, 16480 Stockholm, and also with the
Division of Decision and Control, KTH Royal Institute of Technology, 11428 Stockholm, Sweden. e-mail: gabor.fodor@ericsson.com. }%
\thanks{This work is partially supported by INCT-Signals under grant INCT-25255-82587.32.41/64 (FUNCAP) and 406517/2022-3 (CNPq), and FUNCAP/Universal under grant UNI-0210-00043.01.00/23, and CAPES (Finance code 001). This work is also supported by CNPq under grant 312491/2020-4 and by Ericsson Research, Sweden, and Ericsson Innovation Center, Brazil, under UFC.52 Technical Cooperation Contract Ericsson/UFC.}
}

\maketitle

\markboth{XLIII BRAZILIAN SYMPOSIUM ON TELECOMMUNICATIONS AND SIGNAL PROCESSING - SBrT 2025, SEPTEMBER 29 - OCTOBER 2, 2025, NATAL, RN}{}

%% If writing in English, remove both 'resumo' and 'chave'

%% ---------------------------------------------

\begin{abstract}
This paper introduces an enhanced channel estimation method for reconfigurable intelligent surface (RIS)-assisted multiple-input multiple-output (MIMO) communications. Considering the presence of both the direct link from the base station (BS) to the user equipment (UE) and the indirect link \textit{via} the RIS, we formulate a tensor-based receiver for joint direct and indirect channel estimation called the TenDICE algorithm. The proposed receiver relies on an extended parallel factor analysis (PARAFAC)  tensor model for the received pilot signals, from which the involved (direct and indirect link) channels are estimated using a simple alternating least squares scheme. The proposed TenDICE method is compared with the state-of-the-art least squares (LS) method, Khatri-Rao factorization (KRF) method, enhanced trilinear alternating least squares (E-TALS), and the theoretical Cramér-Rao lower bound (CRLB) based on normalized mean square error (NMSE) and spectral efficiency (SE). The proposed TenDICE outperforms the LS and E-TALS methods and performs similarly to the KRF methods while estimating the direct channel.
\end{abstract}

\renewcommand\baselinestretch{.84}

\section{Introduction}
RIS has been intensely studied over the past few years as a solution to improve wireless communications in terms of SE and coverage \cite{Huang_2019,Gong_2020}. In this context, channel estimation is a significant problem and critical to getting the full benefit of RIS in wireless communications. 

In general and specifically for higher frequencies, the wireless communication propagation medium is subject to attenuation and random fluctuations that compromise and undermine the quality of transmitted information. Due to phenomena such as fading and refraction, the electromagnetic waves responsible for information propagation suffer significant losses that affect the quality of service. As a result, data distortion or even loss of critical information can occur during transmission. Rapid growth in the number of connected devices, particularly in the next generation of wireless communications, i.e., 5G and beyond 5G, has presented significant challenges, especially regarding channel estimation. As a result, extensive research and studies have been conducted in recent years to enhance communication quality and maximize network capacity.

 The passive RIS is a network device capable of engineering and manipulating the characteristics of electromagnetic waves in such a way as to improve the signal-to-noise ratio (SNR) at the UE \cite{BASAR2019}. RIS is crucial in enhancing mobile connectivity and enabling ultra-reliable and massive machine-type communications. Furthermore, RIS-assisted systems show significant potential in addressing challenges such as signal degradation along the transmission path and interference commonly associated with direct line of sight (LoS) connections \cite{LIU2021}.

Tensor-based channel estimation methods enjoy the essential uniqueness properties of tensor decompositions, which are not always covered by matrix estimation methods. The PARAFAC and PARATUCK (hybrid between PARAFAC and TUCKER models) decompositions are examples of tensor decompositions that have been shown to effectively exploit the multidimensional nature of signals in a wireless communication system, including MIMO systems and millimeter-wave communications \cite{DEALMEIDA2007337_2007, 4524041, Gérard_2014, Leandro_2015}, and the design of transceiver devices is more flexible \cite{dearaújo2022semiblindjointchannelsymbol}. According to \cite{10746633}, the communication channel can be modeled as high-order tensors and its geometric structure efficiently exploited in conjunction with tensor decomposition methods.

Several recent works have addressed the problem of channel estimation for RIS-assisted communications using tensor decomposition methods. In \cite{dearaújo2022semiblindjointchannelsymbol}, the authors proposed a MIMO receiver model in a passive RIS-assisted system, where it is possible to estimate the channels semi-blindly jointly and the received data. This work uses the PARATUCK model, whose algebraic structure allows a semi-blind receiver to be developed. The theoretical Cramér-Rao bound (CRB) is also derived to assess channel estimation accuracy. These models commonly adopt an independent and identically distributed (i.i.d.) channel assumption, typically based on the Rayleigh fading channel model. The work  \cite{fazaleasim2023twodimensionalchannelparameterestimation} explores the intrinsic geometrical structure of the channel and also the structure of pilot signals, consequently, decoupling all respective channels (transmitter, RIS, and receiver) that leads to the development of low-complexity methods for channel estimation. %However, the work assumes the direct path between the BS and the UE is blocked. \
However, these numerous works have concentrated on MIMO communication models where only the transmitter-to-RIS and RIS-to-receiver channels are considered, with the direct transmitter-to-receiver path assumed to be blocked. In \cite{dearaújo2022semiblindjointchannelsymbol} the author proposes a two-stage method named as enhanced trilinear alternating least squares (E-TALS) for both direct and indirect channel estimation, where in the first stage, the RIS is turned off to estimate the direct channel, while in the second stage, the indirect channel is estimated \textit{via} bilinear alternating least squares (BALS) by removing the contribution of the direct channel. This two-stage approach has two fundamental shortcomings. First, dividing the direct and indirect (RIS-assisted) channel estimation into two stages implies an extensive processing (and decoding) delay, limiting the performance under channels with shorter coherence times. Second, the sequential estimations of these two links usually involve an error propagation from the first stage to the second stage, affecting the overall system performance. 

Tensor decompositions have been effectively used in various scenarios to develop channel estimation and semi-blind joint detection receivers for MIMO wireless communication systems. In particular, the widespread adoption of the PARAFAC tensor decomposition in wireless communications is attributed to its ability to leverage the inherent multilinear structure of signals and channels, as well as its powerful uniqueness properties \cite{Ximenes_2014, Leandro_2015,Gérard_2014,DEALMEIDA2007337_2007,deAraújo_2020}.

This paper proposes a new tensor-based approach in RIS-assisted communications that can simultaneously estimate the direct and indirect communication links for a downlink MIMO communication scenario. Instead of solving the problems in two stages, we take a different route and formulate a tensor-based received signal model combining direct and indirect link signal contributions. We then recast the resulting received signal tensor as an equivalent third-order PARAFAC tensor model. This structure allows us to derive a novel tensor-based receiver capable of jointly estimating all the involved channels under trivial ambiguities. This is the first work where tensor modeling simultaneously estimates the direct and indirect channels in RIS-assisted MIMO communications. Our solution solves this problem in a single stage, reducing the system latency associated with the channel estimation task, while offering an improved performance over competing schemes.

\subsubsection*{Notations, definitions, and useful properties}
Scalars, vectors, matrices, and tensors are denoted by lower case (\textit{x}), bold lowercase ($\boldsymbol{x}$), bold uppercase $(\boldsymbol{X})$, and calligraphic letter $(\mathcal{X})$, respectively. \textrm{diag}($\boldsymbol{x}$) denotes the diagonal matrix for the $\boldsymbol{x}$ vector along the diagonal. \textrm{vec}($\boldsymbol{X}$) denotes the vectorization operation of the $\boldsymbol{X}$ matrix. \textrm{unvec}($\boldsymbol{x}$) denotes the inverse vectorization operation. $ \boldsymbol{I}_M $ denotes the identity matrix of size $M \times M$. For two matrices $\boldsymbol{A} \in \mathbb{C}^{M \times N}$ and $\boldsymbol{B} \in \mathbb{C}^{P \times N}$, $\boldsymbol{A \diamond B} \in \mathbb{C}^{MP \times N}$ denotes Khatri-Rao product between the $\boldsymbol{A}$ and $\boldsymbol{B}$. $\boldsymbol{A} \odot \boldsymbol{B}$ denotes the Hadamard produtct between $\boldsymbol{A}$ and $\boldsymbol{B}.$ $\boldsymbol{a} \circ \boldsymbol{b}$ is the outer product between $\boldsymbol{a}$ and $\boldsymbol{b}$ vectors. $\boldsymbol{A}^\dagger$ denotes the pseudo inverse of matrix $\boldsymbol{A}$. $\| \boldsymbol{A}\|_F$ denotes the Frobenius norm of $\boldsymbol{A}$ \cite{dealmeida:tel-00460157}.
Throughout this work, we use some important properties, such as \cite{fazaleasim2023twodimensionalchannelparameterestimation}, \cite{dealmeida:tel-00460157}
\begin{equation}\label{eq:1}
    \text{vec}(\boldsymbol{A}\text{diag}(\boldsymbol{b})\boldsymbol{C}^\text{T}) = (\boldsymbol{C \diamond A})\boldsymbol{b}, %\,\,(\text{If} \ \boldsymbol{B} \, \text{is diagonal}),
\end{equation}
\begin{equation}\label{eq:2}
    (\boldsymbol{A \otimes B})(\boldsymbol{C \diamond D}) = (\boldsymbol{AC}) \diamond (\boldsymbol{BD}),
\end{equation}
\begin{equation}\label{eq:3}
    (\boldsymbol{AD_A}) \diamond (\boldsymbol{BD_B}) = (\boldsymbol{A} \diamond \boldsymbol{B})(\boldsymbol{D_A} \odot \boldsymbol{D_B}).
\end{equation}

%\subsection{Tensor preliminaries}
The \textit{n}-mode unfolding of an \textit{N} order tensor $\mathcal{X} \in \mathbb{C}^{I_1 \times I_2 \times I_3 \times ... \times I_N}$ is denoted as $[\mathcal{X}]_n \in \mathbb{C}^{I_n \times I_1I_2 ... I_{n-1}I_{n+1} ... I_N}$. A \textit{N}-order identity tensor is denoted by $\mathcal{I}_{N,R}$. The \textit{n}-mode product between a tensor $\mathcal{X} \in \mathbb{C}^{I_1 \times I_2 \times I_3 \times ... \times I_N}$ with a matrix $\boldsymbol{A} \in \mathbb{C}^{J_n \times I_n}$ is denoted as ($\mathcal{X} \times_n \boldsymbol{A} $) $\in \mathbb{C}^{I_1 \times ... \times I_{n-1} \times J_n \times I_{n+1} \times ... \times I_N}$ \cite{doi:10.1137/07070111X}. 
%\subsection{PARAFAC decomposition}
The PARAFAC decomposition for a tensor $\mathcal{X} \in \mathbb{C} ^{I_1 \times I_2 \times I_3}$ can be represented as $ \mathcal{X} \approx \sum_{r=1}^{R} \boldsymbol{a}_r \circ \boldsymbol{b}_r  \circ \boldsymbol{c}_r$ (where $\boldsymbol{a}_r \in \mathbb{C}^{I_1}, \boldsymbol{b}_r \in \mathbb{C}^{I_2}$  and $\boldsymbol{c}_r \in \mathbb{C}^{I_3}$ are rank-1 tensors components) or $\mathcal{X} \approx \mathcal{I}_{3,R} \times_1 \boldsymbol{A} \times_2 \boldsymbol{B} \times_3 \boldsymbol{C}$ (where $\boldsymbol{A}, \boldsymbol{B}$ and $\boldsymbol{C}$ are the factor matrices) \cite{dealmeida:tel-00460157}, \cite{doi:10.1137/07070111X}.

%\subsection{Some useful properties}

\section{System model}
\begin{figure}[!t] 
\centering
\includegraphics[width=0.4\textwidth]{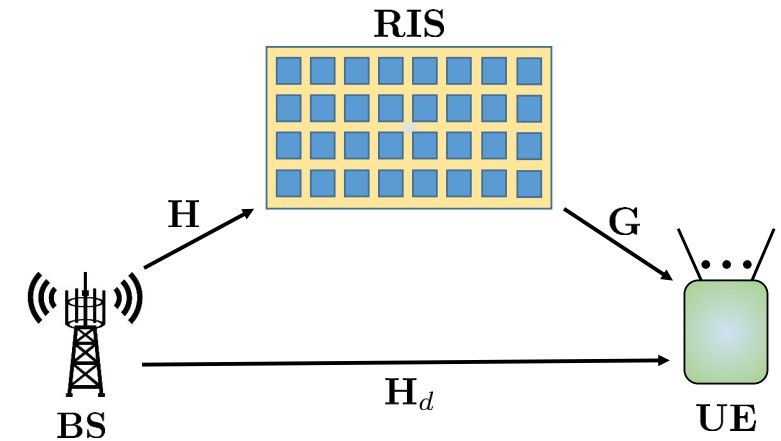}
\caption{RIS-assisted MIMO communication scenario.}
\label{fig:fig1}
\end{figure}

Consider a MIMO communication assisted by an RIS having $N$ reflecting elements. The BS communicates with $M$ antennas to the UE with $Q$ antennas. Furthermore, the BS has a direct channel with the UE $\boldsymbol{H}_d \in \mathbb{C}^{Q \times M}$, as shown in Figure \ref{fig:fig1}. The channel $\boldsymbol{H} \in \mathbb{C}^{N \times M}$ represents the BS-RIS path, while channel $\boldsymbol{G} \in \mathbb{C}^{Q \times N}$ denotes the RIS-UE path. 
The transmitted orthogonal pilot signal $\boldsymbol{S} \in \mathbb{C}^{ M\times T}$, where $T$ consists of the length of the pilot sequence, is received simultaneously \textit{via} both direct and indirect paths. We use Hadamard sequences because they can be transmitted \textit{via} BPSK (Binary Phase Shift Keying) that helps the power amplifier to work close to the saturation region due to the constant modulus property \cite{fazaleasim2023twodimensionalchannelparameterestimation}. For the $p$-th RIS phase-shift pattern, the received signal for both direct and indirect channels is given as:
\begin{equation}\label{eq:4}
    \boldsymbol{X}_p = \left(\boldsymbol{G}\text{diag}\left\{ \boldsymbol{\omega}_p\right\}\boldsymbol{H} + \boldsymbol{H}_d \right)\boldsymbol{S} + \boldsymbol{N}_p \in \mathbb{C}^{Q \times T},
\end{equation}
where $\boldsymbol{\omega}_p$ is the $p$-th phase shift vector of the RIS and $\boldsymbol{N}_p \ \mathtt{\sim} \  \mathcal{CN}(0,\sigma_n^2)$ is the additive white Gaussian noise (AWGN) having $\sigma_n^2$ variance. \eqref{eq:4} can be further written in compact form as:
\begin{equation}\label{eq:5}
    \boldsymbol{X}_p = \boldsymbol{C}\boldsymbol{D}_p\boldsymbol{\bar{H}S}+\boldsymbol{N}_{p} ,
\end{equation}
where the $\boldsymbol{C} = [\boldsymbol{G} \ \boldsymbol{H}_d]   \in \mathbb{C}^{Q \ \times \ (N+M)} $ represents the combined channels, $\boldsymbol{D}_p =  \text{diag}\left\{\boldsymbol{\bar{\omega}}_p \right\} \in \mathbb{C}^{(N + M) \ \times \ (N + M)}$ denotes $p$-th diagonal phase shift matrix, represented as $\boldsymbol{\bar{\omega}} _{p}^\text{T}  = [\boldsymbol{\omega}_p^\text{T} \ \mathbf{1}_M^\text{T}]\in \mathbb{C} ^{1 \ \times \ (N+M)}$ and $\boldsymbol{\bar{H}} = \begin{bmatrix}
    \boldsymbol{H} \\
    \boldsymbol{I}_M
\end{bmatrix}  \in \mathbb{C} ^{ (N + M) \times M}$ is an ``extended'' channel matrix.
Applying matched filtering to the $p$-th received pilot signal can be represented as:
\begin{equation}\label{eq:6}
    \boldsymbol{\bar{Z}}_p = \boldsymbol{X}_p\boldsymbol{S}^{\text{H}} = \boldsymbol{C}\boldsymbol{D}_p\boldsymbol{\bar{H}}+ \boldsymbol{N}_{p}^{'}  \in \mathbb{C}^{Q \times M},
\end{equation}
% Rewriting, 
% \begin{equation}\label{eq:7}
%     \boldsymbol{\bar{Z}}_p = \boldsymbol{C}\boldsymbol{D}_p\boldsymbol{\bar{H}}+,
% \end{equation}
where $\boldsymbol{\bar{Z}}_p \in \mathbb{C}^{Q \times M}$ represents the filtered cascaded channel and $\boldsymbol{N}'_p=\boldsymbol{N}_{p}\boldsymbol{S}^{\text{H}}$ is the filtered noise term. Applying the $\textrm{vec}(\cdot)$ operator to (\ref{eq:6}) and using the property (\ref{eq:1}), we get
\begin{equation}\label{eq:8}
    \text{vec}\boldsymbol{(\bar{Z}}_p) = \text{vec}(\boldsymbol{C}\boldsymbol{D}_p\boldsymbol{\bar{H}})+\text{vec}(\boldsymbol{N}_{p}^{'}),
\end{equation}
\begin{equation}\label{eq:9}
    \boldsymbol{\bar{z}}_p = (\boldsymbol{\bar{H}^\text{T} \diamond C})\boldsymbol{\bar{\omega}}_p +\boldsymbol{n}_{p}^{'} \in \mathbb{C}^{QM \times 1}.
\end{equation}
Equation (\ref{eq:9}) represents the $p$-th filtered signal vector. Collecting all the pilot sequences received and storing them in a matrix
$ \boldsymbol{\bar{Z}} = [\boldsymbol{\bar{z}}_1, \boldsymbol{\bar{z}}_2, \ldots, \boldsymbol{\bar{z}}_P]  \in \mathbb{C}^{QM \times P}$ yields
% \begin{equation}\label{eq:10}
%     \boldsymbol{\bar{Z}} = [\boldsymbol{\bar{z}}_1 , \boldsymbol{\bar{z}}_2, ... , \boldsymbol{\bar{z}}_P]  \in \mathbb{C}^{QM \times P} ,
% \end{equation}
%\eqref{eq:10} can be represented as
\begin{equation}\label{eq:10}
    \boldsymbol{\bar{Z}} = (\boldsymbol{\bar{H}}^\text{T} \diamond \boldsymbol{C}) \boldsymbol{\bar{\Omega}}^\text{T} + \boldsymbol{N{'}} \in \mathbb{C}^{QM \times P}.
\end{equation}
where the equivalent phase-shift matrix is written in compact form as $ \boldsymbol{\bar{\Omega}} = [\boldsymbol{\bar{\omega}}_1, \boldsymbol{\bar{\omega}}_2, \ldots, \boldsymbol{\bar{\omega}}_P]^\text{T}  \in \mathbb{C}^{P \times (N+M) }$. \eqref{eq:10} is the final received signal that contains both direct and indirect channels. %In the next section, \eqref{eq:10} is molded as a third-order tensor to recover the respective channels as detailed below. 

\section{Proposed TenDICE approach}
This section discusses the proposed Tensor-based Direct and Indirect Channel Estimation (TenDICE) algorithm to jointly estimate the equivalent channel matrices  $\boldsymbol{\bar{H}}$ and $\boldsymbol{C}$ containing the information of the direct and indirect links.

\begin{figure}[!t]
\centering
\includegraphics[width=0.35\textwidth]{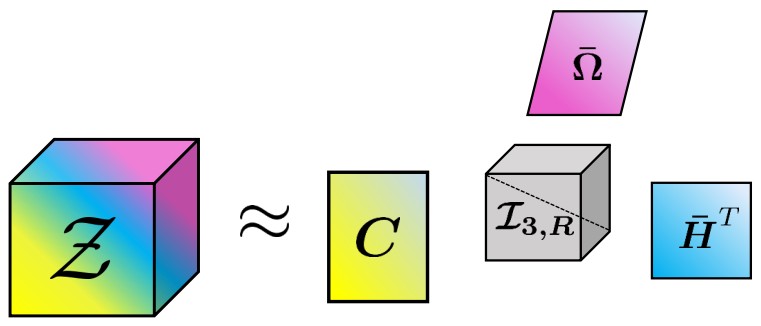}
\caption{Tensor decomposition of the received signal.}
\label{fig:fig2}
\end{figure}

The received signal \eqref{eq:10} can be represented as a third-order tensor model, as follows \cite{9361077}:
\begin{equation}\label{eq:11}
   \boldsymbol{\mathcal{Z}} = \boldsymbol{\mathcal{I}}_{3,(N+M)} \ \times_1 \ \boldsymbol{C} \ \times_2 \ \boldsymbol{\bar{H}}^\text{T} \ \times_3 \ \boldsymbol{\bar{\Omega}} \ \in \mathbb{C}^{Q \times M \times P},
\end{equation}
\begin{equation}\label{eq:12}
    \boldsymbol{\mathcal{\bar{Z}}} = \boldsymbol{\mathcal{Z}} + \boldsymbol{\mathcal{N}} \in \mathbb{C}^{Q \times M \times P},
\end{equation}
where $\boldsymbol{\mathcal{Z}}$ is the noiseless tensor signal (illustrated in Figure \ref{fig:fig2}), $\boldsymbol{\mathcal{I}}_{3,(N+M)}$ is the identity core tensor of third order having rank $R = N+M$, and $\boldsymbol{\mathcal{N}} \in \mathbb{C}^{Q \times M \times P} $ is the third order noise tensor. The received noisy version of tensor $\boldsymbol{\mathcal{Z}}$ can be obtained in \eqref{eq:12}.
In particular, equation (\ref{eq:11}) corresponds to a third-order PARAFAC model for the overall received signal, combining indirect and direct channels. The equivalent channel matrix $\boldsymbol{C}$ can be estimated by solving the following optimization problem \cite{fazaleasim2023twodimensionalchannelparameterestimation}
\begin{equation}\label{eq:13}
   \min_{\boldsymbol{\hat{C}}} \left \| \boldsymbol{[\mathcal{\bar{Z}}]}_{(1)} - \boldsymbol{\hat{C}(\bar{\Omega} \diamond \bar{H}^\text{T})^\text{T} }\right\|  _\text{F}^2,
\end{equation}
where $[\boldsymbol{\mathcal{\bar{Z}}}]_{(1)} \in \mathbb{C}^{Q \times MP}$ is the first mode unfolding of the noisy tensor $\boldsymbol{\mathcal{\bar{Z}}}$ in \eqref{eq:12}, which is given by
\begin{equation}\label{eq:14}
   \boldsymbol{[\mathcal{\bar{Z}}]}_{(1)} = \boldsymbol{C(\bar{\Omega} \diamond \bar{H}^\text{T})^\text{T}} + \boldsymbol{[\mathcal{N}]}_{(1)}.
\end{equation}
Similarly, to estimate $\boldsymbol{\bar{H}}$, the following problem can be solved:
\begin{equation}\label{eq:15}
   \min_{\boldsymbol{\hat{\bar{H}}} }\left \|\boldsymbol{[\mathcal{\bar{Z}}]}_{(2)} - \boldsymbol{\hat{\bar{H}}^\text{T}(\bar{\Omega} \diamond C)^\text{T}}\right\|_\text{F}^2,
\end{equation}
where $[\boldsymbol{\mathcal{\bar{Z}}}]_{(2)} \in \mathbb{C}^{M \times PQ}$ is the second mode unfolding of the noisy tensor $\boldsymbol{\mathcal{\bar{Z}}}$ in \eqref{eq:12}, given by:
\begin{equation}\label{eq:16}
   \boldsymbol{[\mathcal{\bar{Z}}]}_{(2)} = \boldsymbol{\bar{H}}^\text{T} \boldsymbol{(\bar{\Omega} \diamond C)^\text{T}} + \boldsymbol{[\mathcal{N}]}_{(2)},
\end{equation}
The following expressions give the solutions to these problems in the LS sense:
\begin{equation}\label{eq:17}
   \boldsymbol{\hat{C}} = \boldsymbol{[\mathcal{\bar{Z}}]}_{(1)}\boldsymbol{[(\bar{\Omega} \diamond \bar{H}^\text{T})^\text{T}]^\dagger},\,\,
% \end{equation}
% and
% \begin{equation}\label{eq:18}
   \boldsymbol{\hat{\bar{H}}^\text{T}} = \boldsymbol{[\mathcal{\bar{Z}}]}_{(2)} \boldsymbol{[(\bar{\Omega} \diamond C)^\text{T}]^\dagger},
\end{equation}
where the Eq. \eqref{eq:17} denotes the estimates for the combined channels $\boldsymbol{\hat{C}}$ and $\boldsymbol{\hat{H}}$ by LS method

The factor matrices $\boldsymbol{\hat{C}} $ and $\boldsymbol{\hat{\bar{H}}^\text{T}}$ can be estimated in iterative fashion as shown in Algorithm \ref{alg:krf_svd}. The algorithm requires the signal tensor $\boldsymbol{\mathcal{\bar{Z}}}$ \eqref{eq:12} and the output will be the respective equivalent channels as previously described.
\begin{algorithm}[!t]
\caption{TenDICE}
\label{alg:krf_svd}
\begin{algorithmic}[1]
\REQUIRE $\boldsymbol{\mathcal{\bar{Z}}}$ \eqref{eq:12}
\STATE Set $i = 0$
\STATE Set the max number of iterations ($N_{max}$)
\STATE \textbf{Initialize:} $\boldsymbol{\hat{\bar{H}}}$ 
    
    \FOR{$i = 1$ \TO $N_{max}$}
       \STATE $\boldsymbol{\hat{C}}_i \gets \boldsymbol{[\mathcal{\bar{Z}}]}_{(1)}\boldsymbol{[(\hat{\bar{\Omega}} \diamond \hat{\bar{H}}^\text{T})^\text{T}]^\dagger}$
       
        \STATE $\boldsymbol{\hat{\bar{H}}^{\text{T}}}_i \gets \boldsymbol{[\mathcal{\bar{Z}}]}_{(2)}\boldsymbol{[(\hat{\bar{\Omega}} \diamond \hat{C})^\text{T}]^\dagger} $  
     
    \ENDFOR 
    \STATE Compute the estimation error ($\boldsymbol{e}_i$) for $\boldsymbol{[\mathcal{\bar{Z}}]}_{(1)}$
    \STATE Set tolerance for error
    \STATE Do stopping criterion:
    \IF{$i > 1$}
      \IF{$|\boldsymbol{e}_i - \boldsymbol{e}_{i-1}| < $ tol}
          \STATE \textit{Convergence has been achieved.}
          \STATE \textbf{break}
      \ENDIF
    \ENDIF
\STATE \textbf{Return: $\boldsymbol{\hat{C} , \hat{\bar{H}}}$}

\STATE \textbf{End of algorithm}
\end{algorithmic}
\end{algorithm}
\section{Scaling ambiguities and uniqueness}
% {\color{red}During the development of the model, we encountered specific problems related to scaling factors, ambiguities, and uniqueness conditions. Therefore, it was necessary to analyze the number of elements in the system.}
The uniqueness of the involved channel matrices estimates is ensured under mild conditions, thanks to the essential uniqueness properties of the PARAFAC decomposition \cite{doi:10.1137/07070111X,comon:hal-00410057}.
Simple conditions for the identifiability of $\boldsymbol{C}$ and $\boldsymbol{\bar{H}}$ can be obtained from the equations in (\ref{eq:17}) by noting that uniqueness of the LS estimates of these matrices require that $\boldsymbol{\bar{\Omega}} \diamond \boldsymbol{\bar{H}}^\text{T}$ and $\boldsymbol{\bar{\Omega}} \diamond \boldsymbol{C}$ have full column-rank. Otherwise stated, these rank conditions are necessary for the existence of the two right inverses in (\ref{eq:17}) and are translated to
\begin{equation}\label{eq:19}
   P \geq \left(\frac{N + M}{M}\right) \  \text{and }  P \geq \left(\frac{N + M}{Q}\right),
\end{equation}
%where in (\ref{eq:20}) we have the uniqueness conditions for mode-1 and mode-2 unfolds for the tensor $\boldsymbol{\mathcal{Z}}$. 
Combining these conditions, we have 
\begin{equation}\label{eq:20}
   P \geq \text{max}\left(\frac{N + M}{M} , \frac{N + M}{Q}\right).
\end{equation}
% Concerning scaling ambiguities in the factor matrices introduced after the blind decomposition using the ALS method. 
The equivalent channel $\bar{\boldsymbol{H}}$ has an identity matrix in its second block. This structure is exploited at each iteration of ALS. As a result, the channel $\boldsymbol{H}_d$ is estimated without scaling ambiguities. Analyzing the intrinsic structure of the estimated $\hat{\bar{\boldsymbol{H}}}$ and $\hat{\boldsymbol{C}}$ factor matrices including unwanted diagonal scaling ambiguities as $\boldsymbol{D_{\bar{H}}} \in \mathbb{C}^{(N+M) \times (N+M)}$ and $\boldsymbol{D_C} \in \mathbb{C}^{(N+M) \times (N+M)}$, respectively, given as 
\begin{equation}\label{eq:21}
\boldsymbol{\hat{\bar{H}}^{\text{T}}} = 
\underbrace{
\begin{bmatrix}
     & \boldsymbol{D_{H}} & &\\
     & & \boldsymbol{I}_M & \\
\end{bmatrix}
}_{\boldsymbol{D_{\bar{H}}}}
\begin{bmatrix}
    \boldsymbol{H} \\
    \boldsymbol{I}_M
\end{bmatrix},
\end{equation}
and
\begin{equation}\label{eq:22}
\boldsymbol{\hat{C}} = 
\begin{bmatrix}
    \boldsymbol{G} \ \boldsymbol{H}_d
\end{bmatrix}
\underbrace{
\begin{bmatrix}
    & \boldsymbol{D_G} & &\\ 
     & & \boldsymbol{D_{H_D}} & \\
\end{bmatrix}.
}_{\boldsymbol{D_C} }
\end{equation}
To build the final estimate of the cascaded channel $\boldsymbol{\hat{T}}=\hat{\boldsymbol{H}}^\text{T} \diamond \hat{\boldsymbol{G}}$, we use the property \eqref{eq:3} to obtain
$\boldsymbol{\hat{T}}=\boldsymbol{\hat{H}}^{\text{T}} \diamond \hat{\boldsymbol{G}} = (\boldsymbol{D}_\text{H}\boldsymbol{H}^{\text{T}})^{\text{T}} \diamond (\boldsymbol{G}\boldsymbol{D}_G) = (\boldsymbol{\hat{H}^{\text{T}}} \diamond \hat{\boldsymbol{G}})\boldsymbol{I}_{N}$, since $\boldsymbol{D}_\text{H}=(\boldsymbol{D}_G)^{-1}$.
% \begin{equation}\label{eq:23}
%  \boldsymbol{\hat{H}}^{\text{T}} \diamond \hat{\boldsymbol{G}} = (\boldsymbol{D}_\text{H}\boldsymbol{H}^{\text{T}})^{\text{T}} \diamond (\boldsymbol{G}\boldsymbol{D}_G) = (\boldsymbol{\hat{H}^{\text{T}}} \diamond \hat{\boldsymbol{G}})\boldsymbol{I}_{N}.
% \end{equation}
Hence, we conclude that the estimated cascaded channel does not need scaling factor correction, since the scaling ambiguities cancel each other. Note that the direct channel estimate $\hat{\boldsymbol{H}}_d$ already comes without scaling ambiguity since we exploit the structure of $\boldsymbol{\bar{H}}$ by enforcing the second block to be the identity matrix in each ALS iteration (see (\ref{eq:21})). This implies that $\boldsymbol{D_{H_D}}$ in (\ref{eq:22}) is equal to an identity matrix.

\section{Simulation results and discussion}
In this section, we present numerical results for evaluating the performance of the proposed TenDICE algorithm and comparing it with state-of-the-art competing methods. The performance metrics include the normalized mean squared error (NMSE) for different channel estimation methods and the spectral efficiency (SE). The NMSE is calculated as
\begin{equation}\label{eq:24}
\text{NMSE}(\boldsymbol{\hat{T}}) = \frac{1}{R} \sum\limits_{r = 1}^R \frac{\|\boldsymbol{T}^{(r)} - \boldsymbol{\hat{T}}^{(r)}\|_F^2}{\|\boldsymbol{T}^{(r)}\|_F^2},
\end{equation}
\begin{equation}\label{eq:25}
\text{NMSE}(\boldsymbol{\hat{H}}_d) = \frac{1}{R} \sum\limits_{r = 1}^R \frac{\|\boldsymbol{H}_d^{(r)} - \boldsymbol{\hat{H}}_d^{(r)}\|_F^2}{\|\boldsymbol{H}_d^{(r)}\|_F^2},
\end{equation}
where $\boldsymbol{\hat{T}}^{(r)} = \boldsymbol{\hat{H}}_{(r)}^{\text{T}} \boldsymbol{\diamond \ \hat{G}}_{(r)}  $ , and $\boldsymbol{\hat{H}}_d^{(r)}$ denotes the $r$-th run, and $R$ is the total number of Monte Carlo iterations. Calculating SE involves the design of the precoder $\boldsymbol{f} \in \mathbb{C}^{M \times 1}$ and the combiner $\boldsymbol{w} \in \mathbb{C}^{Q \times 1}$ based on recovered channels \cite{fazaleasim2023twodimensionalchannelparameterestimation}. According to the authors, the algorithm involves two stages, where the first is based on least squares Khatri-Rao factorization (LSKRF) in the cascaded channel, and the second involves singular value decomposition (SVD). Note that the SE is only based on the cascaded channel $\hat{\boldsymbol{H}}^\text{T} \diamond \hat{\boldsymbol{G}}$, only. Considering a RIS with optimal phases vector ($\boldsymbol{\omega}_{\text{opt}}$), and noise variance $\sigma_n^2$, the SE is calculated as
\begin{equation}\label{eq:26}
\text{SE = log}_{2} \left(\left |1 + \frac{| \boldsymbol{w}^\text{H}\boldsymbol{\hat{G}}\text{diag}(\boldsymbol{\omega}_{\text{opt}})\boldsymbol{\hat{H}f}|^2}{ \sigma_n^2 }\right | \right).
\end{equation}
Both metrics are being measured as a function of signal-to-noise ratio (SNR), in dB, defined as $\text{SNR(dB) = 10log}_{10} \left(\frac{P_T}{ \sigma_n^2 } \right)$,
% \begin{equation}\label{eq:27}
% \text{SNR(dB) = 10log}_{10} \left(\frac{P_T}{ \sigma_n^2 } \right),
% \end{equation}
where $P_T$ is the power transmission (in W).

\begin{figure}[!t]
\centering
\includegraphics[width=0.4\textwidth]{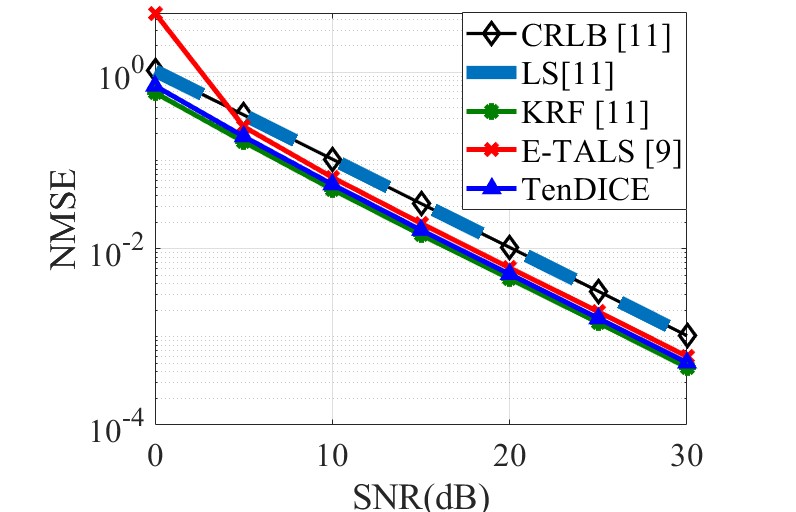}
\caption{TenDICE performance (estimation of the cascaded channel $\boldsymbol{T}$).}
\label{fig:fig3}
\end{figure}

\begin{figure}[!t]
\centering
\includegraphics[width=0.4\textwidth]{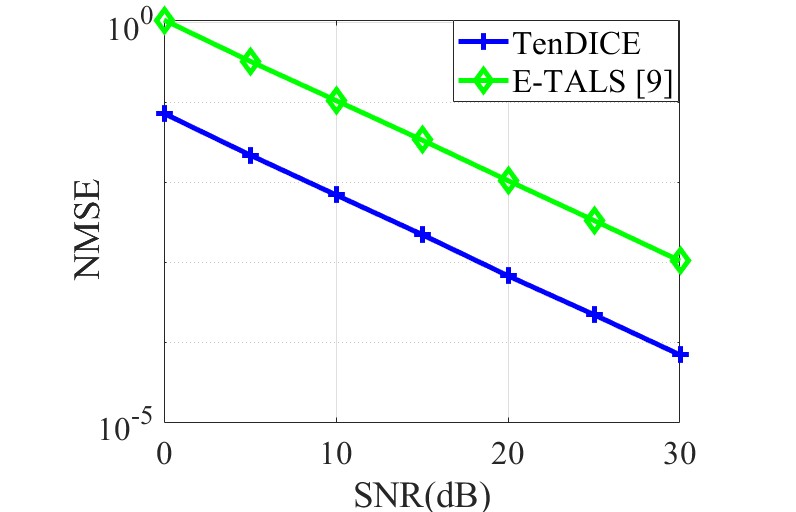}
\caption{TenDICE performance (estimation of the direct channel $\boldsymbol{H}_d$).}
\label{fig:fig5}
\end{figure}

\begin{figure}[!t]
\centering
\includegraphics[width=0.4\textwidth]{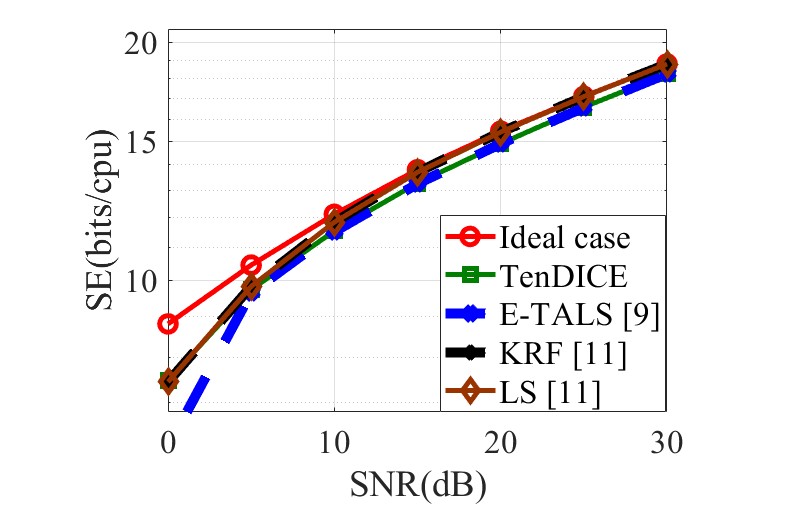}
\caption{SE performance (in bits/bits per channel use) (RIS-assisted link).}
\label{fig:fig4}
\end{figure}

In Fig. \ref{fig:fig3}, the proposed method TenDICE is compared to the other competing methods, i.e., the LS method \cite{9361077}, the KRF method \cite{9361077}, unstructured CRLB \cite{9361077}, and the two-stage method E-TALS \cite{dearaújo2022semiblindjointchannelsymbol}. We consider $N=16$ reflecting elements at the RIS, $M=4$ number of antenna elements at the BS, and $Q=4$ antenna elements at the UE, respectively. The proposed TenDICE method outperforms the classical LS \cite{9361077} and two-stage E-TALS \cite{dearaújo2022semiblindjointchannelsymbol} methods, and maintains similar performance as the KRF method \cite{9361077}. However, unlike the competing methods, our proposed TenDICE method still estimates the direct channel simultaneously. The E-TALS method \cite{dearaújo2022semiblindjointchannelsymbol} estimates the direct channel and, in the second stage, estimates the indirect channel by employing the direct channel from the first stage. Consequently, the error propagates from the first stage to the second stage, deteriorating the overall performance. Secondly, the two-stage approach is also spectrally inefficient and has an estimation delay for getting all the indirect and direct channels. Fig. \ref{fig:fig5} shows the NMSE base performance comparison of the proposed TenDICE method and the first stage of the \cite{dearaújo2022semiblindjointchannelsymbol} E-TALS method, where matched filtering is employed to estimate the direct channel. It is shown that our proposed TenDICE method substantially outperforms the competing method, proving that exploiting the tenosr structure improves the noise rejection and thus improves the performance. 

 Fig. \ref{fig:fig4} shows the SE performance of the proposed method compared with the competing LS, KRF, and E-TALS methods. To have a fair comparison, we also assumed the precoder and combiner on true channels. It is worth mentioning that the LS and the KRF methods show similar performance as compared to the proposed TenDICE method, while E-TALS shows a slight degradation in the lower SNR regime. All the methods show identical performance as compared to the ideal case in the higher SNR regime. %but the SE for him was just a little bit bigger than the others (it is essential to mention that we tested several values of $N$ and noticed that the SE had an increase, showing that for a RIS with a large number of elements the proposed model may be efficient). 
Analyzing Table II, we can conclude that as the number of RIS elements increases, the execution time for each method also increases. Still, TenDICE has a much higher execution time than the other methods, but with the benefit of simultaneous direct and indirect channels. 
\begin{center}
\begin{table}[!t]
\label{tab:tab1} \caption{Simulation parameters} 
    \centering
    \begin{tabular} { |c|c| } 
 \hline
   Parameter & Value   \\
 \hline
 $N$ & $\{8,16,32,64,128\}$    \\ 
 \hline
 $P$ (time slots) & $N$    \\ 
 \hline
 $M$ & 4   \\ 
 \hline
 $Q$ & 4  \\ 
 \hline
 $T$ & $M$  \\ 
 \hline
 MC (Monte Carlo) & $10^3$  \\ 
 \hline
 Tol (tolerance) & $10^{-6}$  \\ 
 \hline
 $P_T$ (Tx power) &   1   \\
 \hline
\end{tabular}
\end{table}
\end{center}

\begin{center}
\begin{table}[!t]
\label{tab:tab2} \caption{Runtime for different values of $N$ in each method} 
    \centering
    \begin{tabular} { |c|c|c|c| } 
 \hline
   $N$ & LS   & KRF &  TenDICE  \\
 \hline
 8 & 0,0648 ms  &  0,457 ms & 0,968 ms     \\ 
 \hline
 16 & 0,2 ms & 1,3 ms &     1,7 ms    \\ 
 \hline
 32 & 0,716 ms & 3,1 ms &    6,2 ms    \\ 
 \hline
 64 & 1,5 ms & 4,5 ms &   12,6  ms  \\ 
 \hline
 128 & 12,4 ms & 14,4 ms &     108,2 ms   \\ 
 \hline
\end{tabular}
\end{table}
 \label{eq:table1}
\end{center}

 \vspace{-3.5ex}
 
\section{Conclusions and Outlook}
This paper proposed an estimation algorithm for direct and indirect channels simultaneously in a passive RIS communication system. The TenDICE model has shown promise for this type of estimation, according to the results of NMSE and SE, compared to LS, KRF, the two-stage E-TALS methods, and the CRLB as the reference bound. At this point, it is essential to mention that we are using unstructured channels and therefore LS satisfies CRLB \cite{fazaleasim2023twodimensionalchannelparameterestimation}. Another important thing about our proposed model is that we do not need to correct the scale factor when recovering the direct channel, and this is due to the matrix and tensor properties that are features of the model and the effect of the direct channel that removes scaling ambiguities. Alternative RIS configurations can be explored for future work. For example, in \cite{dealmeida2024channelestimationdiagonalris}, the authors proposed tensor-based channel estimation schemes for beyond diagonal (BD)-RIS, which implies non-diagonal scattering matrices. In this context, extending the TenDICE receiver to BD-RIS is a topic for future work. Considering RIS with hardware impairments \cite{10137372} and evaluating its impact on the proposed receiver's performance is also a topic under study. Furthermore, for future work, the computational complexity of TenDICE can be explored and compared with other estimation methods.  %In this case, the direct channel can be used for compensation and even improve the system performance in the presence of these impairments.

%\section{Acknowledgements}
% The authors thank the partial support of FUNCAP under grant ITR-0214-00041.01.00/23, and the National Institute of Science and Technology (INCT-Signals) sponsored by Brazil's National Council for Scientific and Technological Development (CNPq) under grant 406517/2022-3. This work is also supported by CNPq under grant 312491/2020-4 and by Ericsson Research, Sweden, and Ericsson Innovation Center, Brazil, under UFC.52
% Technical Cooperation Contract Ericsson/UFC.

\renewcommand\baselinestretch{.9}

\bibliographystyle{IEEEtran}
\bibliography{References}

\end{document}